\documentclass[12pt,final]{elsart}

  \usepackage{latexsym,bm,amsmath,amssymb,amsfonts}
  \usepackage{epsfig,graphics,graphicx}
  \usepackage{slashed}

  \long\def\comment#1{ }

  \newcommand{\beq}{\begin{eqnarray}}
  \newcommand{\eeq}{\end{eqnarray}}
  
 \def\simge{\mathrel{%
   \rlap{\raise 0.511ex \hbox{$>$}}{\lower 0.511ex \hbox{$\sim$}}}}
\def\simle{\mathrel{
   \rlap{\raise 0.511ex \hbox{$<$}}{\lower 0.511ex \hbox{$\sim$}}}}

\begin{document}

\begin{frontmatter}

\parbox[]{16.0cm}{ \begin{center}
\title{Nonequilibrium 2PI evolution of the $O(N)$ model \\ with longitudinal expansion}

\author{Yoshitaka Hatta$^{a}$  and Akihiro Nishiyama$^{b}$ }

\address{$^{a}$ Faculty of Pure and Applied Sciences, University
of Tsukuba, Tsukuba, Ibaraki 305-8571, Japan
}

\address{$^{b}$ Maskawa Institute for Science and Culture, Kyoto Sangyo University, Kyoto 603-8555, Japan}


\end{center}

\begin{abstract}
Motivated by the problem of thermalization in heavy--ion collisions, we present numerical simulations of the nonequilibrium evolution of the $O(N)$ model in $1+2$ dimensions with longitudinal expansion and in the presence of a background field. We work in the NLO approximation of the $1/N$ expansion and consider both the strong and weak coupling cases. Special emphasis is put on the difference between the 2PI formalism and the classical statistical approach. In the 2PI case at  strong coupling, we find some evidence of the Bose--Einstein (exponential) distribution in the expanding system.
\end{abstract}
}

\end{frontmatter}

\section{Introduction}

Understanding how and when thermalization is achieved in the fireball created in heavy--ion collision experiments is one of the most pressing, but difficult questions in physics of the Quark Gluon Plasma (QGP). The system starts out with a very peculiar initial condition described by the Color Glass Condensate (CGC) framework \cite{Kovner:1995ts}, and eventually ends up with the hydrodynamic regime characterized by the thermodynamic properties of QCD. The transient regime in between is a `missing link' about which our knowledge is severely limited due to the highly nonlinear dynamics of QCD which is furthermore complicated by the one--dimensional expansion of the system.

Analytical insights are available for certain aspects of the evolution, and various scenarios for thermalization have been put forward. [See \cite{Kurkela:2011ti} and references therein.] But if one wants to go beyond parametric estimates and get the whole evolution under quantitative control,  numerical simulations based on the first principles seem to be  necessary.
 So far, simulations based on the classical statistical approximation have proven to be very practical for this purpose \cite{Romatschke:2005pm,Berges:2007re,Kunihiro:2010tg,Dusling:2011rz,Berges:2012ev}. These simulations nicely demonstrate, among others, the exponential growth  of soft gluons triggered by the decay of  classical fields. However, they also indicate that in the weak coupling picture the unstable regime develops in an unrealistically (from the viewpoint of  heavy--ion phenomenology) large span of time. Besides, the classical statistical approximation is a low--momentum effective theory, so `thermalization' achieved in this framework is  the classical Rayleigh--Jeans type, and  not the quantum Bose--Einstein distribution or the Boltzmann (exponential) distribution which is often regarded as a signal of the QGP in heavy--ion collisions. It is then difficult to tell whether one can form the truly equilibrated QGP on a reasonable time scale, or just some far--from--equilibrium gluonic matter which mimics the QGP by acquiring  a constant energy--pressure relation (`equation of state') due to the phenomenon of prethermalization \cite{Berges:2004ce}.

 In principle, the latter difficulty can be solved by the two--particle irreducible (2PI) formalism  which has a variety of applications in nonequilibrium physics and has been recently discussed in the context of heavy--ion collisions in the CGC picture \cite{Hatta:2011ky} (see, also, \cite{Nishiyama:2010mn}). In the 2PI formalism, the Bose--Einstein distribution is guaranteed as it consistently includes the relevant terms responsible for the quantum equilibration. However, unlike the classical statistical approach, usually the collision terms in the 2PI approach are evaluated to fixed order in the coupling constant. While such an approximation may be justified for certain purposes, in the presence of instability at weak coupling it will break down in the soft sector because higher order loop diagrams, proportional to large powers of the particle density, are no longer suppressed at the height of instability.

 Given that the classical statistical approach and the 2PI formalism have their own advantages (and disadvantages), one may hope to find a unified approach which captures both these  advantages in a complementary manner. Unfortunately such a formalism is not available in QCD at the moment.
There is, however, one (and only one) known example where this can be done---It is the scalar $O(N)$ theory at large $N$ \cite{Aarts:2002dj}. To next--to--leading order (NLO) in the $1/N$ expansion, one can resum the 2PI loop diagrams (including the terms omitted in the classical statistical approach) to all orders in the coupling constant. It is thus a very attractive toy model from which one can learn important lessons for QCD. Numerical simulations of this model have been previously performed   in the flat coordinates \cite{Aarts:2001yn,Arrizabalaga:2004iw}. In this paper we report the result of numerical simulations of this model in (1+2)--dimensions with one--dimensional expansion (one time dimension, one transverse dimension, and one longitudinal (expanding) dimension)  and in the presence of a background field. Our main focuses are the difference between the 2PI and the classical statistical approaches, and  the impact of longitudinal expansion on the equilibration process. For  previous 2PI simulations in  (spherically) expanding systems, see  \cite{Tranberg:2008ae}.

\section{The model}
Our model field theory is the $O(N)$--symmetric $\lambda \phi^4$ theory in (1+2)--dimensions $x^\mu=(t,x_T,z)$
\beq
S= \int dt dx_T dz \left(\frac{1}{2} \partial_\mu \phi_a \partial^\mu \phi_a -\frac{1}{2}m^2 \phi_a \phi_a -\frac{\lambda \sigma}{4!N}(\phi_a \phi_a)^2 \right)\,,
\eeq
 where $a=1,2,\cdots,N$.
 $\lambda$ is the dimensionless coupling constant and the dimensionful parameter $\sigma$ sets the energy scale of the problem.
  We assume that the system is undergoing one--dimensional expansion in the $z$--direction. It is then convenient to switch to the `$\tau$--$\eta$'  coordinates defined by
 \beq
 \tau=\sqrt{t^2-z^2}\,, \qquad \eta = \tanh^{-1} \frac{z}{t}\,.
  \eeq
   $\tau$ is the so--called proper time, and $\eta$ is the  rapidity.
 In these coordinates, the action reads
\beq
S= \int \tau d\tau dx_T d\eta  \left(\frac{1}{2} (\partial_\tau \phi_a)^2 -\frac{1}{2\tau^2}(\partial_\eta \phi_a)^2 -\frac{1}{2} (\partial_{T} \phi_a)^2 -\frac{1}{2}m^2 \phi_a \phi_a -\frac{\lambda \sigma}{4!N}(\phi_a \phi_a)^2 \right)\,,  \label{ver}
\eeq
and the classical equation of motion takes the form
\beq
\left(\partial_\tau^2
+ \frac{1}{\tau} \partial_\tau
-\frac{1}{\tau^2} \partial_\eta^2
- \partial_T^2 + m^2 + \frac{\lambda \sigma}{6N}\phi_b\phi_b \right) \phi_a =0\,.
\eeq
 We shall be interested in the evolution of the system   characterized by a nonzero field expectation value at the initial time $\tau=\tau_0 \approx 0$
 \beq
\frac{\langle \phi_a(\tau_0) \rangle}{\sqrt{\sigma}}= \delta_{a1}\sqrt{\frac{6N}{\lambda}}\,,
\label{clas}
\eeq
 which we take to be boost invariant (independent of $\eta$) and, for simplicity, homogeneous (independent of $x_T$).
   Such a background field is to mimic the initial condition of heavy--ion collisions in the color glass condensate picture. In that case, weak (QCD) coupling is assumed so that the classical gluon fields are strong $A^\mu \sim {\mathcal O}(1/\sqrt{\alpha_s}) \gg 1$. Here we relax this assumption and treat $\lambda$ as a free parameter.

 At $\tau>\tau_0$ the classical field decays, and fuels energy for quantum fluctuations to grow. The latter is characterized by the statistical function
\beq
F_{ab}(x,x') = \frac{1}{2}\langle \{\phi_a(x), \phi_b(x')\}\rangle- \langle \phi_a(x)\rangle \langle \phi_b(x')\rangle\,,
\eeq
 and the spectral function
\beq
\rho_{ab}(x,x')=i\langle [\phi_a(x),\phi_b(x')]\rangle\,.
\eeq
We choose the initial condition such that $\langle \phi_a(\tau)\rangle=0$ for $a\neq 1$ at all times, and introduce the $O(N-1)$--symmetric notation $F_{11}\equiv F_{\parallel}\,$, $F_{22}=F_{33}=\cdots =F_{NN}\equiv F_\perp$.
Since the classical field is independent of $x_T$ and $\eta$, we work in the Fourier space and denote the conjugate momenta to $x_T$ and $\eta$ by  $p_T$ and $p_\eta$, respectively.

 The evolution of the system is described by the following set of coupled equations (`Kadanoff--Baym equation') for $\langle \phi \rangle$, $F$ and $\rho\,$:
 \beq
 \left[ \partial_\tau^2
+ \frac{1}{\tau} \partial_\tau
 + m^2 + \frac{\lambda \sigma}{6N}\left( \phi^2(\tau) + 3F_{11}(\tau,\tau)+\sum_{b\neq 1}F_{bb}(\tau,\tau)  \right) \right]\phi(\tau) \nonumber \\
   = -\int^\tau_{\tau_0} \tau' d\tau'\Sigma^\rho_{11}(\tau,\tau')\phi(\tau')\,, \label{1}
\eeq
where
\beq
F(\tau,\tau) \equiv \int \frac{dp_T dp_\eta}{(2\pi)^2} F(\tau,\tau,p_T,p_\eta)\,,
\label{tad}
\eeq
and
$\langle \phi_1(\tau)\rangle$ is simply denoted as $\phi(\tau)$.
The equations for the statistical and the spectral functions are
\beq
&& \left[\left(\partial_\tau^2 + \frac{1}{\tau}\partial_\tau +\frac{p_\eta^2}{\tau^2}+ p_T^2 \right)\delta_{ab}
+ M_{ab}^2(\phi) \right] F_{bc}(\tau,\tau',p) \nonumber \\ && = -\int^\tau_{\tau_0} \tau''d\tau'' \Sigma_{ab}^\rho(\tau,\tau'',p)F_{bc}(\tau'',\tau',p) + \int^{\tau'}_{\tau_0} \tau''d\tau'' \Sigma^F_{ab}(\tau,\tau'',p)\rho_{bc}(\tau'',\tau',p)\,, \label{2}
\eeq
\beq
\left[\left(\partial_\tau^2 + \frac{1}{\tau}\partial_\tau +\frac{p_\eta^2}{\tau^2}+ p_T^2\right)\delta_{ab}
+M_{ab}^2(\phi) \right] \rho_{bc}(\tau,\tau',p) = -\int^\tau_{\tau'} \tau'' d\tau'' \Sigma^\rho_{ab}(\tau,\tau'',p)\rho_{bc}(\tau'',\tau',p)\,, \nonumber \\  \label{3}
\eeq
  where the effective mass term is
 \beq
 M_{ab}^2(\phi)=m^2 \delta_{ab}+  \frac{\lambda \sigma}{6N} \left( F_{cc}(\tau,\tau)+\phi^2\right)\delta_{ab} +
 \frac{\lambda \sigma}{3N}\left( F_{ab}(\tau,\tau)+\phi^2\delta_{a1}\delta_{b1}\right)\,.  \label{mass}
 \eeq
 The self energies $\Sigma^F_{ab}$ and $\Sigma^\rho_{ab}$ are evaluated in the 2PI formalism to  next--to--leading order (NLO) approximation in the $1/N$ expansion. Their expressions are rather lengthy and are not  reproduced here. We refer the reader to the original paper \cite{Aarts:2002dj} and, importantly for the present purpose, its extension \cite{Berges:2010nk} to systems with a certain class of curved coordinates including the $\tau$--$\eta$ coordinates.

\section{Numerical simulations: Setup}

Thanks to the manifestly causal structure, the equations (\ref{1})--(\ref{3}) can be straightforwardly solved on a lattice.
 We consider a lattice of size $64\times 64$ in the momentum space with the periodic boundary condition. The momentum and time are discretized as
\beq
\frac{p_T}{\sigma} = \frac{2\pi n_T}{64 a_T \sigma} = \frac{\pi n_T}{64}, \qquad  p_\eta = \frac{2\pi n_\eta}{64 a_\eta} = \frac{\pi n_\eta}{16}\,, \qquad -32 \le n_T, n_\eta \le 32\,,
\eeq
\beq
 \tau = na_t\,, \qquad \tau_0 = 10 a_t\,,
\eeq
where we have chosen the lattice spacings $a_T=2/\sigma$, $a_\eta=0.5$, and $a_t=0.4/\sigma$.
Due to the symmetry of the problem $p \leftrightarrow -p$, the actual simulations may be performed on a $32 \times 32$ lattice by restricting to the region $p_\eta, p_T>0$. As for the mass parameter, we take $m=0.1 \sigma$.\footnote{The bare mass $m$ is negligible almost everywhere when solving the evolution equation. The only reason we keep $m$ finite is to avoid an IR divergence in the integral (\ref{ren}) in our renormalization prescription,  $\int dp_T F_0 \sim \sum_{n_T} F_0$ at $n_T=0$.}

The initial conditions are determined as follows.  In the noninteracting theory, and in the expanding coordinates, the statistical function takes the generic form (see Appendix)
\beq
 F_0(\tau,\tau',p) = \frac{\pi}{2}Re \Bigl\{H^{(2)}_{ip_\eta}(m_T \tau) H^{(1)}_{ip_\eta}(m_T \tau') \Bigr\} \left(\frac{1}{2}+n_p\right)\,, \label{in}
 \eeq
 where\footnote{In actual simulations, the transverse momenta squared is replaced by
\beq
\frac{p_T^2}{\sigma^2} \to \frac{2}{(a_T\sigma)^2}\left(1-\cos \frac{\pi n_T}{32}\right)
=\frac{1}{2} \left(1-\cos \frac{\pi n_T}{32}\right)\,,
\eeq
in order to reduce lattice artifacts.} $m_T=\sqrt{p_T^2+m^2}$ and
 \beq
n_p \delta^{(2)}(p-q)= \langle a_p^\dagger a_q\rangle\,,
 \eeq
  is an arbitrary distribution of pre-existing quanta. We assume that the initial state contains only the classical field (\ref{clas}) and  no on--shell particle excitations. This means  $n_p(\tau_0)=0$, and the classical field induces an effective mass term for the fluctuation. Our initial conditions are thus given by
\beq
F_{\parallel} (\tau,\tau',p) &=&\frac{\pi}{4}Re \Bigl\{H^{(2)}_{ip_\eta}(m_\parallel \tau) H^{(1)}_{ip_\eta}(m_\parallel \tau') \Bigr\}\,, \qquad  m_\parallel= \sqrt{p_T^2+m^2+3\sigma^2}\,,   \nonumber \\
F_\perp (\tau,\tau',p) &=& \frac{\pi}{4}Re \Bigl\{H^{(2)}_{ip_\eta}(m_\perp \tau) H^{(1)}_{ip_\eta}(m_\perp \tau') \Bigr\} \,,  \qquad
 m_\perp = \sqrt{p_T^2 + m^2 + \sigma^2} \,, \label{init}
\eeq
evaluated at $\tau=\tau'=\tau_0$.

Since the equations (\ref{1})--(\ref{3}) are second order in the $\tau$--derivative, we must also specify  the first derivatives. For the classical field, we take
\beq
\partial_\tau \phi(\tau)\big\arrowvert_{\tau_0}= 0\,, \qquad (\tau_0\to 0)\,, \label{initial}
\eeq
  again motivated by the situation in the CGC case
\cite{Kovner:1995ts}, whereas $\partial_\tau F(\tau,\tau_0)|_{\tau_0}$ and
$\partial_\tau \partial_{\tau'}F(\tau,\tau')|_{\tau_0}$ are computed from
(\ref{init}). The initial condition for the spectral function is fixed by the canonical commutation relation
\beq
\rho_{ab}(\tau_0,\tau_0,p)=0\,, \qquad \partial_\tau \rho_{ab}(\tau,\tau_0)\big\arrowvert_{\tau=\tau_0} = \delta_{ab}\frac{1}{\tau_0}\,. \label{rhorho}
\eeq
 (\ref{rhorho}) is actually satisfied for all values of $\tau$ even in the interacting theory, so it can be used as a check of the simulation.

 Concerning renormalization, in (1+2)--dimensions the tadpole self--energy (\ref{tad})
 is linearly divergent. We renormalize this by subtracting  the free part at each step of the evolution
\beq
F(\tau,\tau) \to \int \frac{d^2p}{(2\pi)^2} \left(F(\tau,\tau,p)- F_0(\tau,\tau,p) \right)\,,
 \label{ren} \eeq
where $\int F_0$ is computed from (\ref{in}) (with $n_p=0$). There is also a logarithmic divergence in the `sunset' diagram which appears in $\Sigma$ at two loops. Though in principle this requires an extra term in the mass  renormalization (see, e.g., \cite{Juchem:2003bi}),  in this work we do not perform the renormalization of logarithmic divergences.

As stated in the introduction, our main interest is the difference between the full 2PI calculation and its classical statistical approximation.  We thus show the result of the latter simulation as well.  In the present context (at NLO in $1/N$), this simply amounts to neglecting, in the set of equations (\ref{1})--(\ref{3}),  terms quadratic in $\rho$'s when they appear in conjunction with terms quadratic in $F$'s \cite{Aarts:2001yn}. Note that we still perform the renormalization (\ref{ren}), so in this sense our `classical' simulation is not exactly the same as the other  classical statistical simulations  where cut--off dependent tadpoles are kept as they are.

Furthermore, in order to see the effect of the expansion, we also perform simulations in the non-expanding case in the flat coordinates $(t,x_T,z)$ on an isotropic lattice $a_z = 2/\sigma = a_T$, $a_t=a_\tau$. The parameters are chosen such that  the relation
 $a_z = \tau_0 a_\eta$ (from $dz = \tau d\eta$
  at midrapidity $\eta=z=0$) is satisfied  at the initial time $t_0=\tau_0$.
 The initial conditions for the classical field are the same as in the expanding case (up to the replacement $\tau \to t$), while those for the $F$, $\rho$--functions are
 \beq
 F_{\parallel}(t_0,t_0,p)&=&\frac{1}{2\sqrt{p^2+m^2+3\sigma^2}}\,, \qquad (p^2=p_T^2+p_z^2)\,, \nonumber \\
 F_\perp(t_0,t_0,p) &=&  \frac{1}{2\sqrt{p^2+m^2+\sigma^2}}\,,
 \eeq
 \beq
  \rho_{ab}(t,t,p)=0\,, \qquad \partial_t \rho_{ab}(t,t',p)\big\arrowvert_{t=t'} =\delta_{ab}\,, \label{freerho}
  \eeq
 and $\partial_t F(t,t_0,p)\arrowvert_{t=t_0}=0$, etc. (c.f., (\ref{free})).

The degree of thermalization may be inferred from the functional form of $F$ at large time. In the non-expanding case, we extract the effective distribution $n_p(t)$ 
from the data using the prescription
\beq
\frac{1}{2}+ n_p(t) \equiv \sqrt{F(t,t',p)\partial_t \partial_{t'} F(t,t',p)}\big\arrowvert_{t=t'}\,,
\label{freean}
\eeq
 which is often employed in the literature (e.g., Refs.~\cite{Aarts:2001yn,Tranberg:2008ae}).
The `equilibrium' distribution in the expanding case is not known, nor is it clear  if such a distribution is theoretically well--defined in our anisotropically  expanding system. We try, however, the following ansatz  for the late--time behavior of the correlation function (cf. Eqs.~(\ref{in}) and (\ref{asy}))
\beq
F(\tau,\tau,p) = \frac{C}{\omega_p \,\tau} \left(\frac{1}{2} + \frac{1}{e^{\omega_p/T}-1} \right)\,, \qquad \omega_p = \sqrt{p_T^2+m_{eff}^2}\,, \label{ansatz}
\eeq
  and see if we can identify an exponentially decaying (in $p_T$) component by adjusting the parameter $C(\tau)\approx 1$ (`field strength renormalization'). At the same time we determine the fitting parameters $T(\tau)$ (`temperature') and $m_{eff}(\tau)$ (`thermal effective mass').

  Note that, as already implied by (\ref{ansatz}), we neglect the dependence of $F$ on $p_\eta$, or equivalently, the longitudinal momentum $p_z \sim p_\eta/\tau$ at late times.  Indeed, in our simulations $F$ becomes independent of $p_\eta$ after some time because $p_\eta$ enters the equation of motion in the form $(p_\eta/\tau)^2$ (see, also, (\ref{correction})). In order to see the dependence on $p_z\sim p_\eta/\tau$, we have to use a considerably larger $p_\eta$ cutoff such that the typical value of $p_\eta/\tau$ at large $\tau$ is of the same order as the $p_T$ cutoff.
  We do not perform such simulations in this paper both from technical and phenomenological reasons. Technically, 2PI simulations in a very large (and anisotropic) lattice is computationally very expensive.  Besides, the product of Hankel functions used in the initial condition (\ref{init}) behaves badly when $p_\eta$ becomes larger than around 10. [Our current cutoff is $p_\eta <2\pi$.] Phenomenologically, in heavy--ion collisions, experimentalists usually measure the $p_T$ dependence, but not the $p_z$ dependence (to our knowledge) because the latter is not boost invariant and depends sensitively on the collision energy. Our primary goal here is to reproduce the exponential $p_T$ dependence around midrapidity $\eta\approx 0$ which is well established experimentally, and for this purpose the inclusion of finite--$p_z$ modes is presumably not very crucial: Due the the redshift $p_z \sim p_\eta/\tau$, the relevant values of $p_z$ naturally decrease with time,  and modes with finite $p_z$ escape from the midrapidity region and do not affect much the $p_T$ distribution there.

\section{Numerical simulations: Results }

In this section we present the numerical solution of the equations (\ref{1})--(\ref{3}) obtained after $1600$ steps  $(\tau/\tau_0=160)$ of evolution in time.  We run $2^3=8$ different simulations corresponding to the following cases: (i)  weak $\lambda=10^{-4}$ and strong   $\lambda =10$ coupling, (ii) 2PI and the classical statistical approximation, (iii) with and without  longitudinal expansion. All the results below are for $N=4\,$.

\subsection{Weak coupling}

\begin{figure}[h]
\begin{tabular}{c}
\includegraphics[height=6cm]{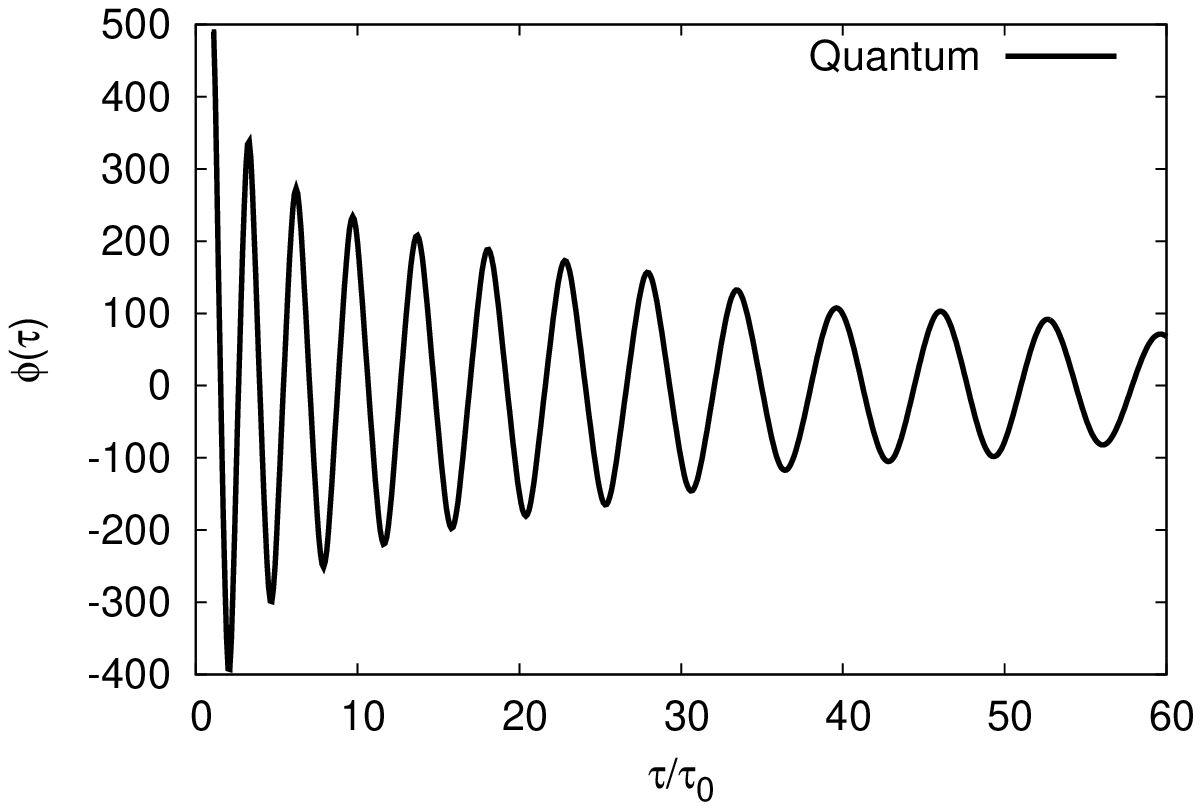}
\end{tabular}
\begin{tabular}{c}
\includegraphics[height=6cm]{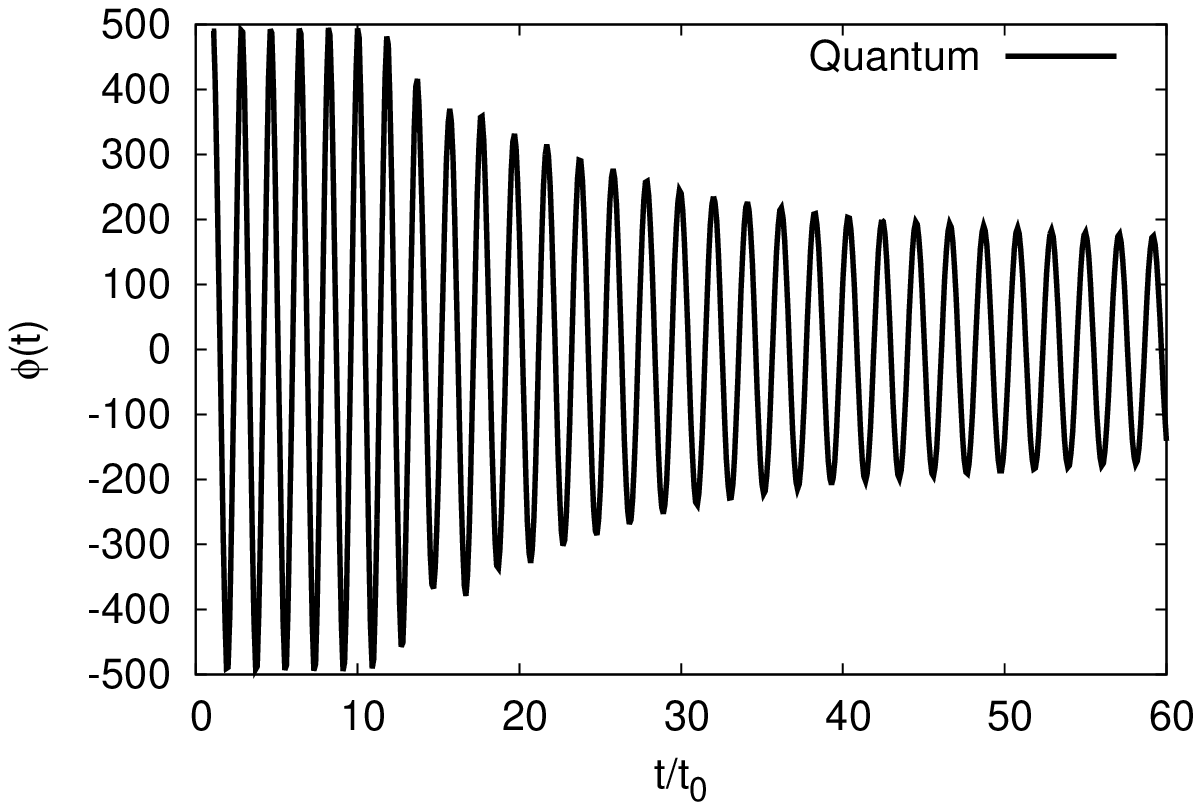}
\end{tabular}
\caption{Time evolution of the classical field at weak coupling $\lambda = 10^{-4}$. Left: with expansion; Right: without expansion.
\label{phi}}
\end{figure}

Let us first present the results at weak coupling $\lambda = 10^{-4}$.
Fig.~\ref{phi} shows the time evolution of the classical field (in units of $\sqrt{\sigma}$) in the expanding case (left) and non-expanding case (right).
 In the expanding case, the amplitude of $\phi$ decays very fast and has a characteristic shape. This can be understood analytically \cite{Berges:2012iw} at weak coupling where the evolution of $\phi$ is largely determined by the classical equation. A detailed analysis gives $\phi(\tau) \sim 1/\tau^{1/3}$, while the period of oscillation is constant in $\tau^{2/3}$, meaning that the effective period becomes longer at larger $\tau$. These tendencies are clearly visible. In the non-expanding case,  on the other hand, the amplitude is constant in time at the classical level, and diminishes only as a result of its coupling to quantum fluctuations. The difference between the full 2PI and the classical statistical approximation is  unnoticeable, hence we did not plot the latter.

\begin{figure}[h]
\begin{tabular}{c}
\includegraphics[height=6cm]{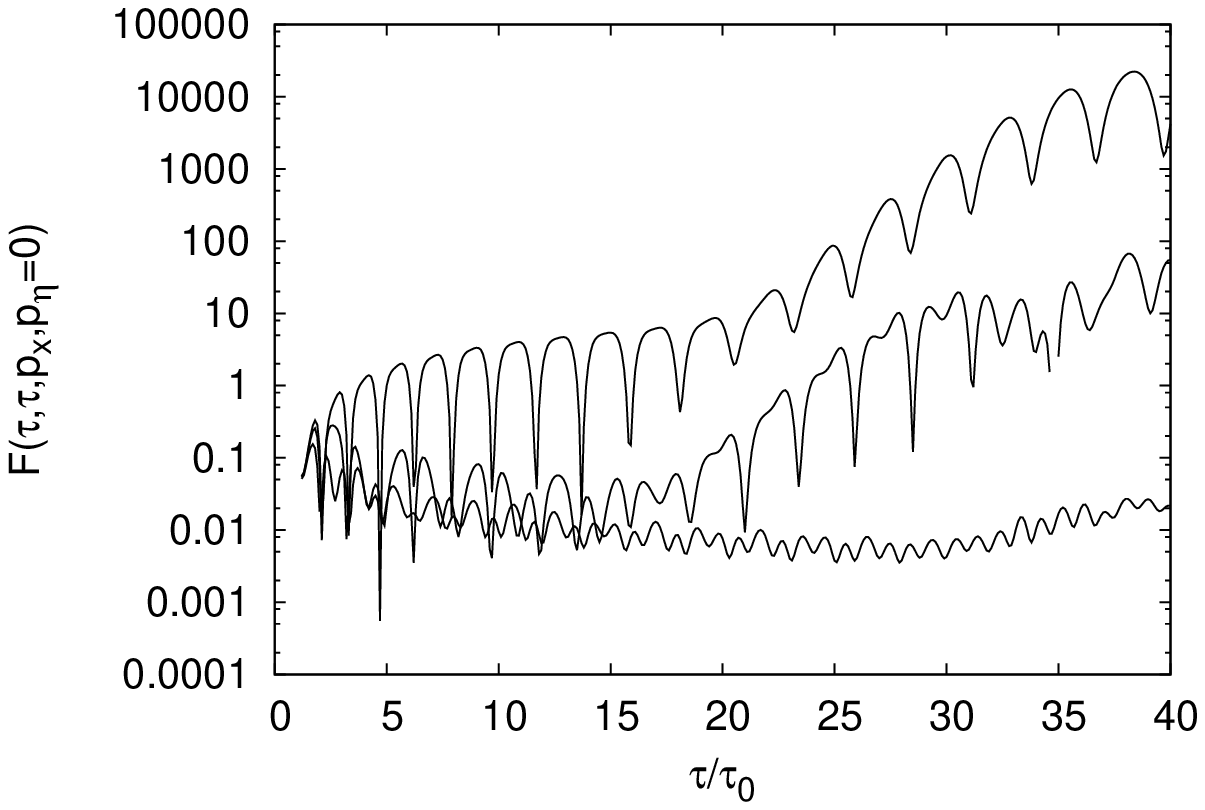}
\end{tabular}
\begin{tabular}{c}
\includegraphics[height=6cm]{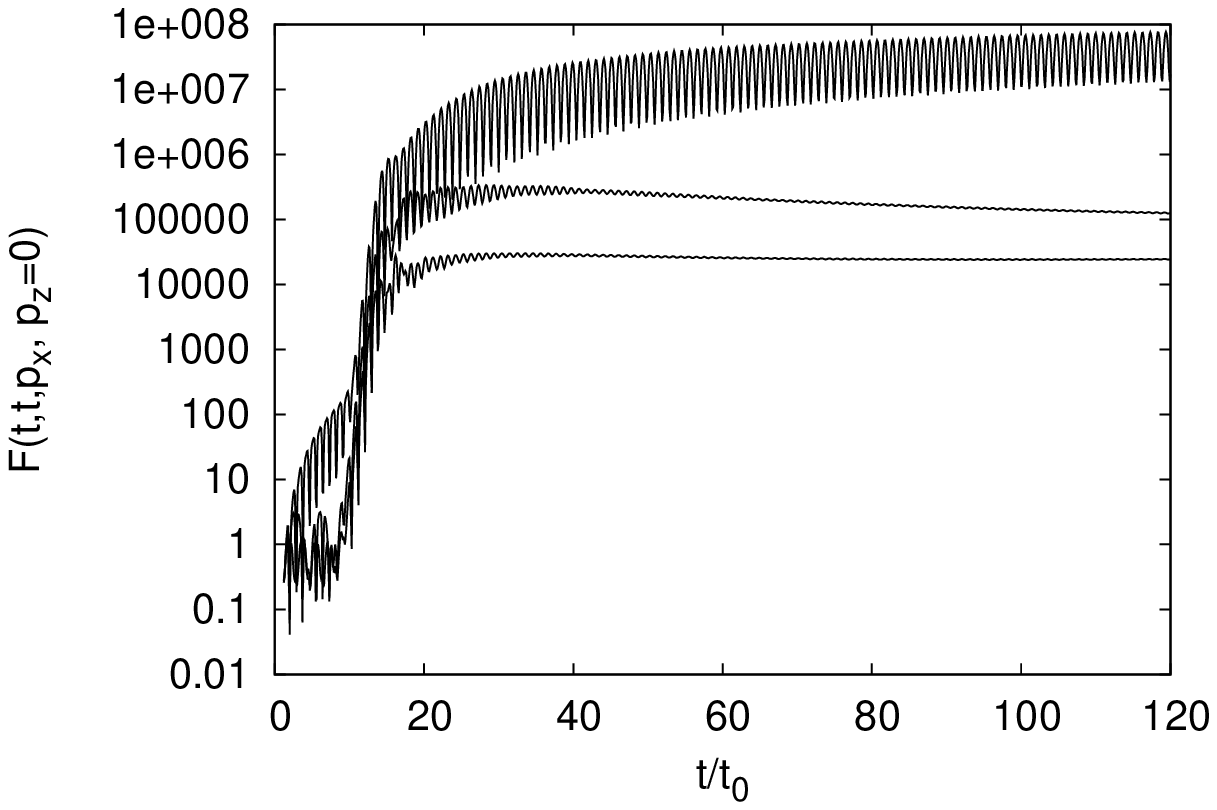}
\end{tabular}
\caption{Time evolution of the statistical function at weak coupling $\lambda = 10^{-4}$ for three different values of $n_T$: $n_T=0,8, 16$ (from top to bottom). Left: with expansion. Right: without expansion.
\label{F1}}
\end{figure}

 Next, in Fig.\ref{F1} we plot $F_{\parallel}(\tau,\tau,p_T,p_\eta=0)$ (expanding case, left) and $F_{\parallel}(t,t,p_T,p_z=0)$ (non-expanding case, right)\footnote{In the nonexpanding case, we actually plot $\, \sigma F(t,t,p)$ to make it dimensionless.}   for three different values of $p_T/\sigma=\frac{\pi   }{64}n_T\,$, $n_T=0,8,16$. We confirm the parametric resonance previously observed at weak coupling \cite{Berges:2012iw}.  We also see that the effect of the expansion is huge, suppressing the magnitude of $F$ by several orders of magnitude  compared with the non-expanding case. Actually,  the suppression is so strong that the equal--time correlator $F(\tau,\tau,p)$, which should be a positive quantity, turns slightly negative in some momentum modes beyond $ \tau \gtrsim 50\tau_0$.  We presume this is because of large errors incurred when very large and very small numbers (which differ by $\sim 10^6$) coexist, or simply due to an artifact of the smallness of our lattice.\footnote{This problem does not seem to occur in the Monte Carlo simulation (rather than solving differential equations) of the classical statistical theory on a larger lattice \cite{Berges:2012iw}. It is computationally very expensive and not realistic for us to increase the lattice size. } We note, however, that $F(\tau,\tau,p)$ turns positive again for all momentum modes at later times $\tau \gtrsim 100\tau_0$, and shows a smooth, nonthermal distribution in $p_T$.

 Again, there is no discernible difference between the 2PI and the classical statistical approximation. This is understandable in the non-expanding case as $F$ becomes very large and the condition $FF \gg \rho\rho$ is always satisfied. [Remember that $\rho$ is an ${\mathcal O}(1)$ quantity due to the constraint (\ref{freerho}).] In the expanding case, it is simply because the whole collision term is small and contributes very little to the evolution. In either case, the lack of any difference in the two simulations indicates that the system is far from the true equilibrium state even at the final time $\tau/\tau_0=160$. The $p_T$--distribution is so overpopulated in the soft region that it is entirely impossible to find any hint of `subtle' quantum effects like the exponential distribution. Presumably we have to wait for a much longer time to see the difference, but this is beyond the scope of the present work.

\subsection{Strong coupling}

\begin{figure}[h]
\begin{tabular}{c}
\includegraphics[height=6cm]{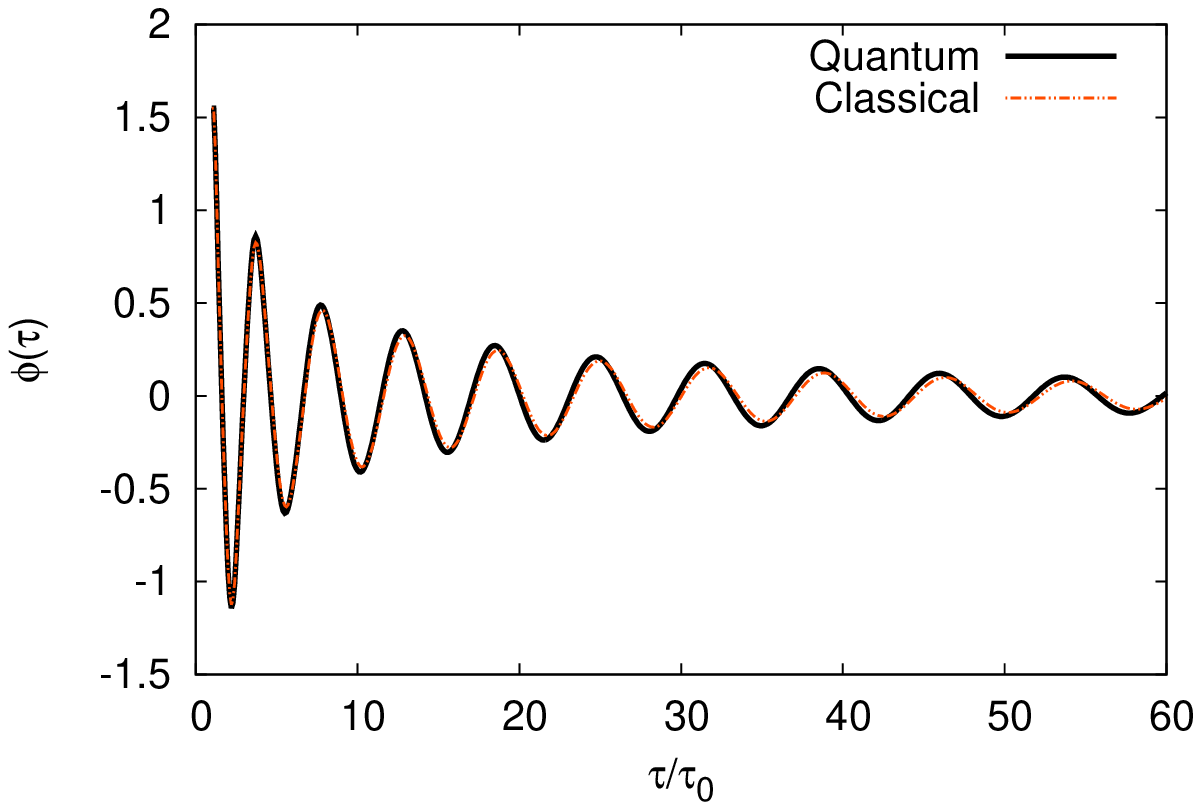}
\end{tabular}
\begin{tabular}{c}
\includegraphics[height=6cm]{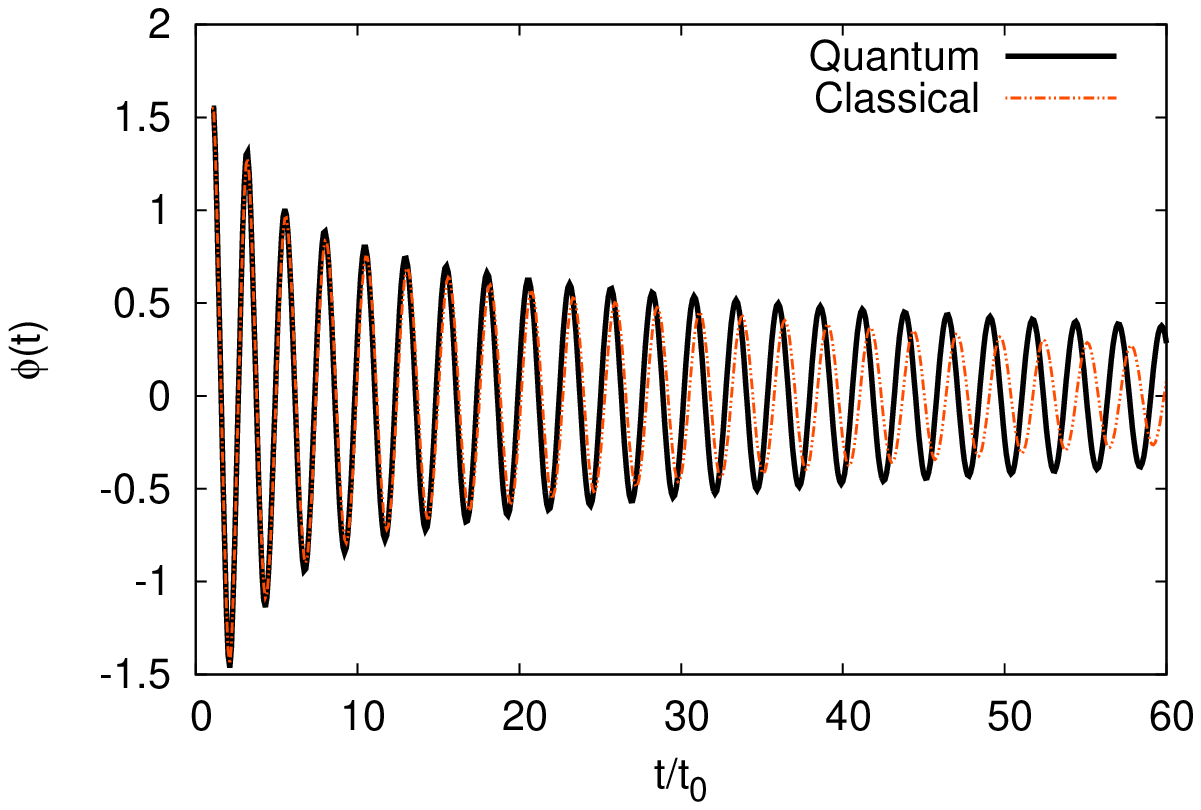}
\end{tabular}
\caption{Time evolution of the classical field at strong coupling $\lambda = 10$. Left: with expansion; Right: without expansion. Both the 2PI result (black solid line) and the classical statistical result (red dashed line) are shown, although the difference is very small.
\label{phi2}}
\end{figure}

We now turn to the  results at strong coupling $\lambda = 10$. In Fig.~\ref{phi2} we plot the time evolution of the classical field. Compared to the weak coupling case (Fig.~\ref{phi}), the decay of the amplitude is faster because the transfer of energy to quantum fluctuations is more efficient at strong coupling. Still, the difference between the 2PI and classical simulations is barely noticeable.

\begin{figure}[h]
\begin{tabular}{c}
\includegraphics[height=6cm]{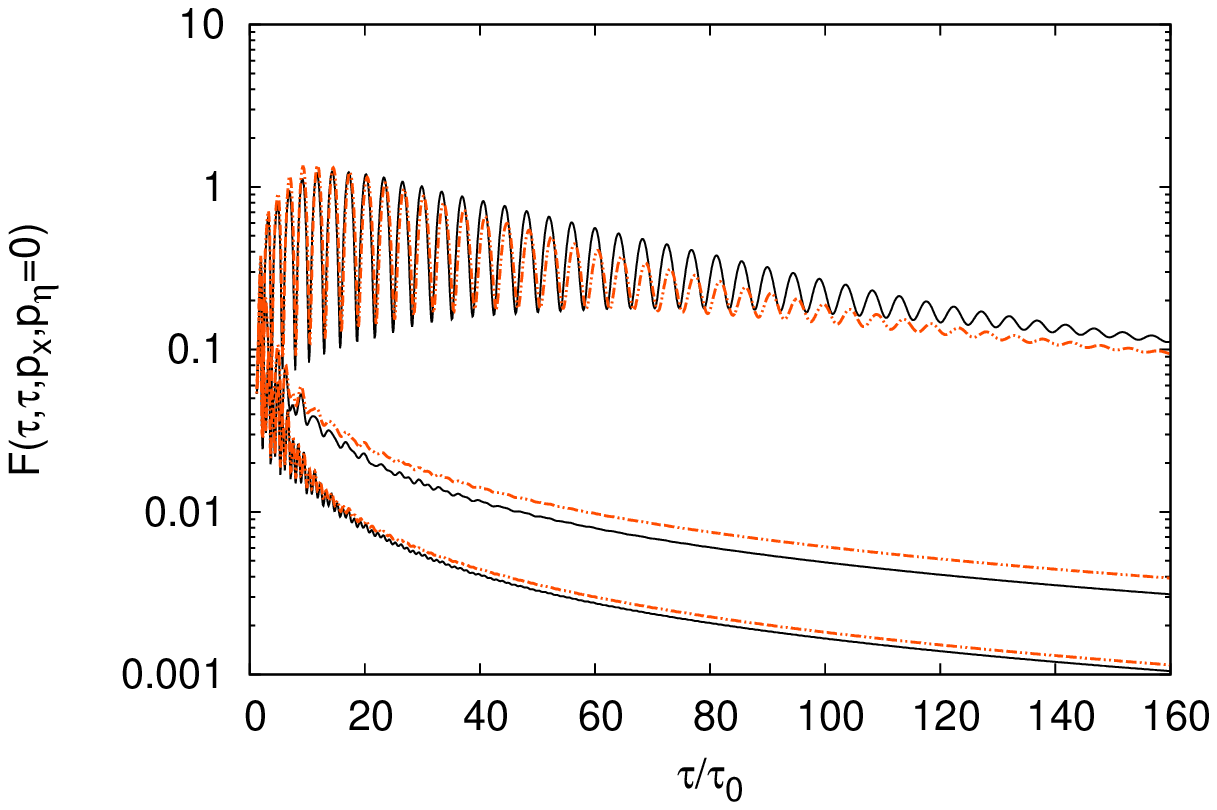}
\end{tabular}
\begin{tabular}{c}
\includegraphics[height=6cm]{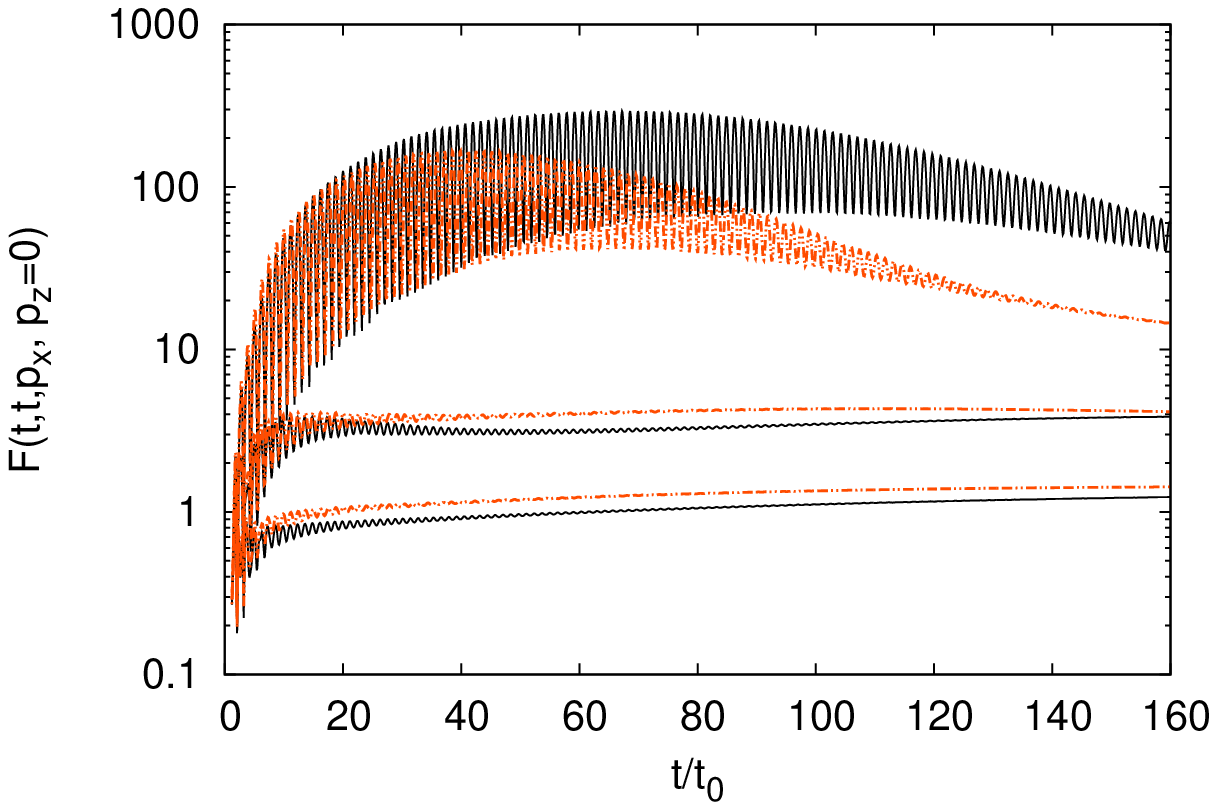}
\end{tabular}
\caption{Time evolution of the statistical function at strong coupling $\lambda = 10$ for three different values of $n_T$: $n_T=0$, $n_T=8$, $n_T=16$ (from top to bottom). Left: with expansion. Right: without expansion.
\label{F2}}
\end{figure}

\begin{figure}[h]
\begin{tabular}{c}
\includegraphics[height=6cm]{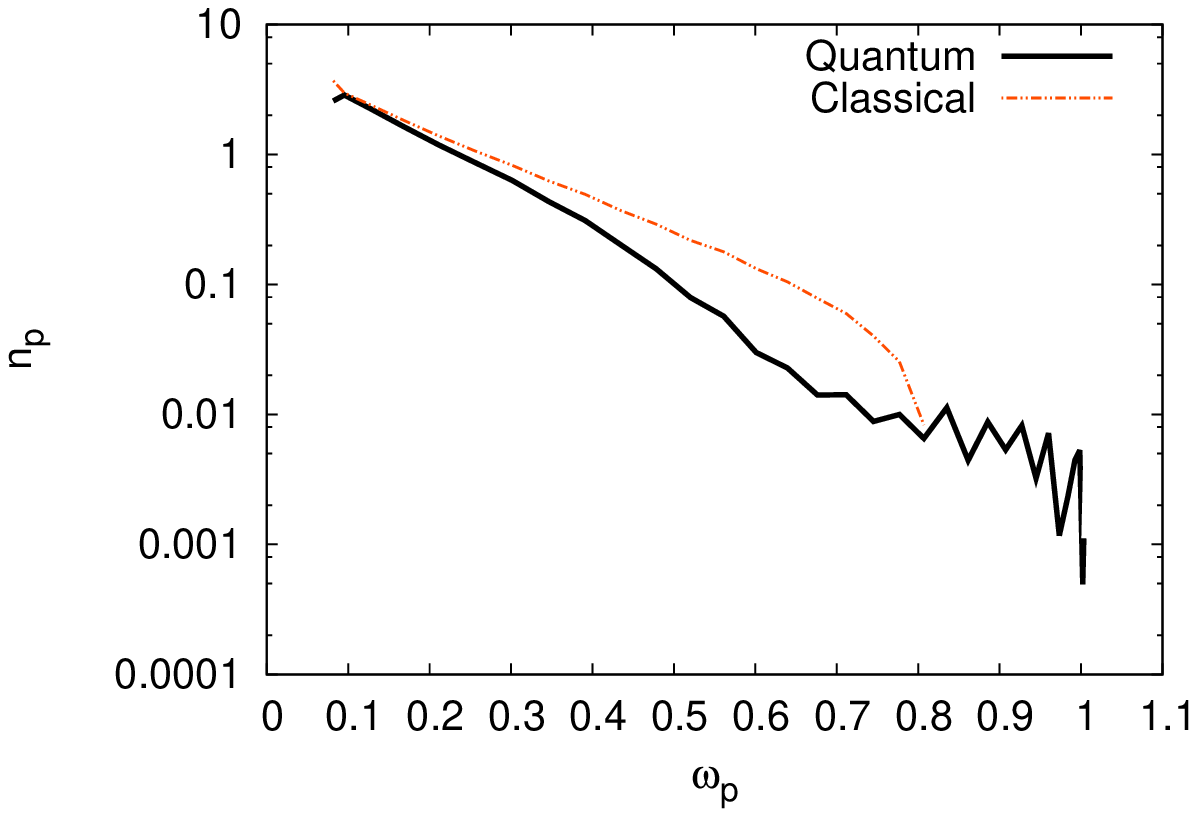}
\end{tabular}
\begin{tabular}{c}
\includegraphics[height=6cm]{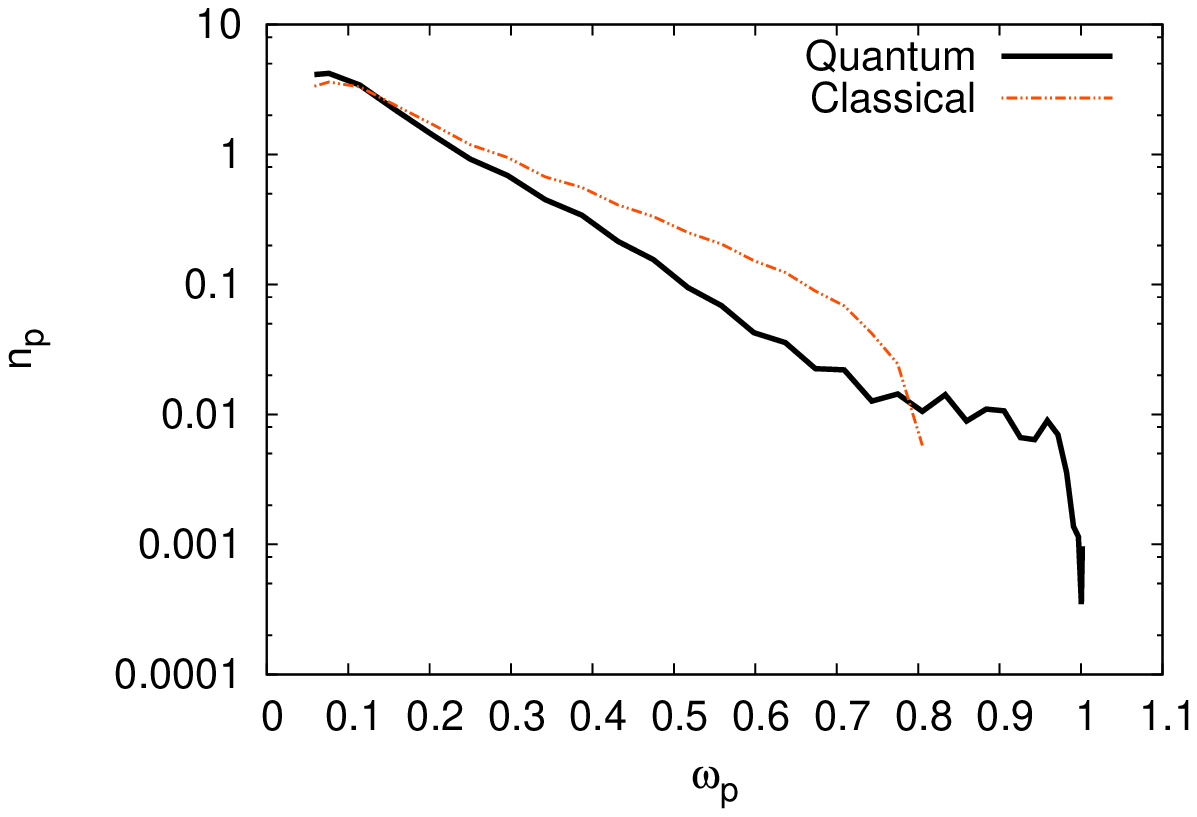}
\end{tabular}
\caption{The distribution $n_p$ as a function of $\omega_p$ at strong coupling $\lambda = 10$  in the expanding case. Left: $\tau=50\tau_0$. Right: $\tau=150\tau_0$. The distribution from the classical statistical approximation (red line) turns negative beyond $\omega_p\gtrsim 0.8$.
\label{n1}}
\end{figure}

Next we plot the statistical function at strong coupling in Fig.~\ref{F2} which should be compared  with the weak coupling counterpart Fig.~\ref{F1}.
 Because the magnitude of the classical field is smaller,  the parametric resonance is less pronounced, and is basically unrecognizable in the expanding case. At least in the present model, this ineffectiveness of the instability appears to be the key to achieve an early  thermalization.
 Indeed, we get a reasonable fit of the momentum distribution using the ansatz (\ref{ansatz}).
 Fig.~\ref{n1} shows the distribution $n_p(\tau)\equiv \tau \omega_p  F(\tau,\tau,p)/C-\frac{1}{2}\,$ in the expanding case at two values of $\tau\, $: $\ \tau/\tau_0=50$ and $\tau/\tau_0=150$.   The fitting parameters are found to be
\beq
\begin{tabular}{|l|l|c|r|}
\hline
  & $C$ & $T/\sigma$ & $m_{eff}/\sigma$ \\ \hline
$\tau/\tau_0=50$ & 0.905 & 0.354 & 0.0817 \\ \hline
$\tau/\tau_0=150$ & 0.913 & 0.413 & 0.0586 \\
\hline
\end{tabular}
\nonumber
\eeq
  where $C\approx 1$, as expected.  The exponential behavior actually  starts to show up already around $\tau/\tau_0 \gtrsim 40$.  The difference between the 2PI and the classical statistical approximation is now manifest. In the latter case (red lines), the distribution plunges
and becomes negative in the high--$p_T$ region\footnote{This is an expected behavior because the equilibrium distribution in the classical statistical theory corresponds to the following approximation to the Bose--Einstein distribution
\beq
\frac{1}{e^{\omega_p/T}-1} \to \frac{T}{\omega_p}-\frac{1}{2}\,.
\eeq } (around $n_T =15\sim 20$) where the quantum (2PI) distribution  still keeps up with the exponential trend.

\begin{figure}[h]
\begin{tabular}{c}
\includegraphics[height=6cm]{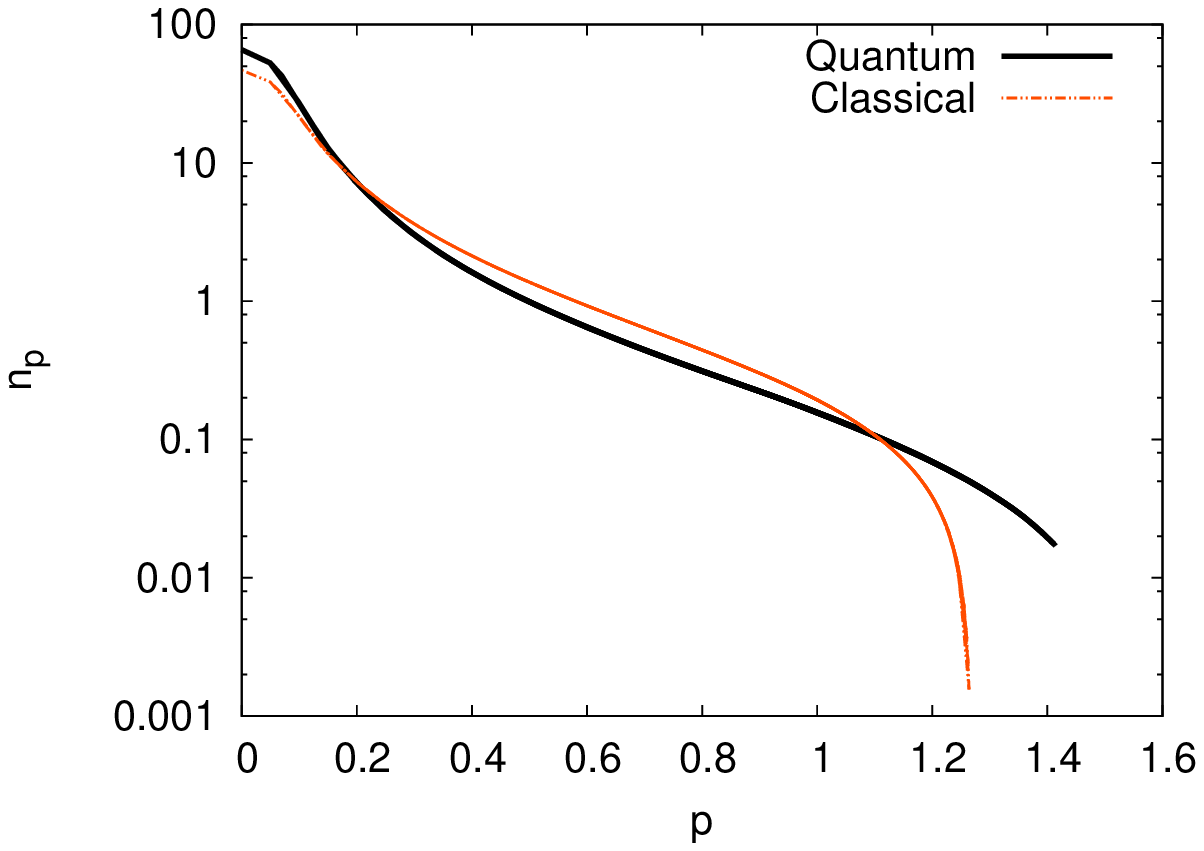}
\end{tabular}
\begin{tabular}{c}
\includegraphics[height=6cm]{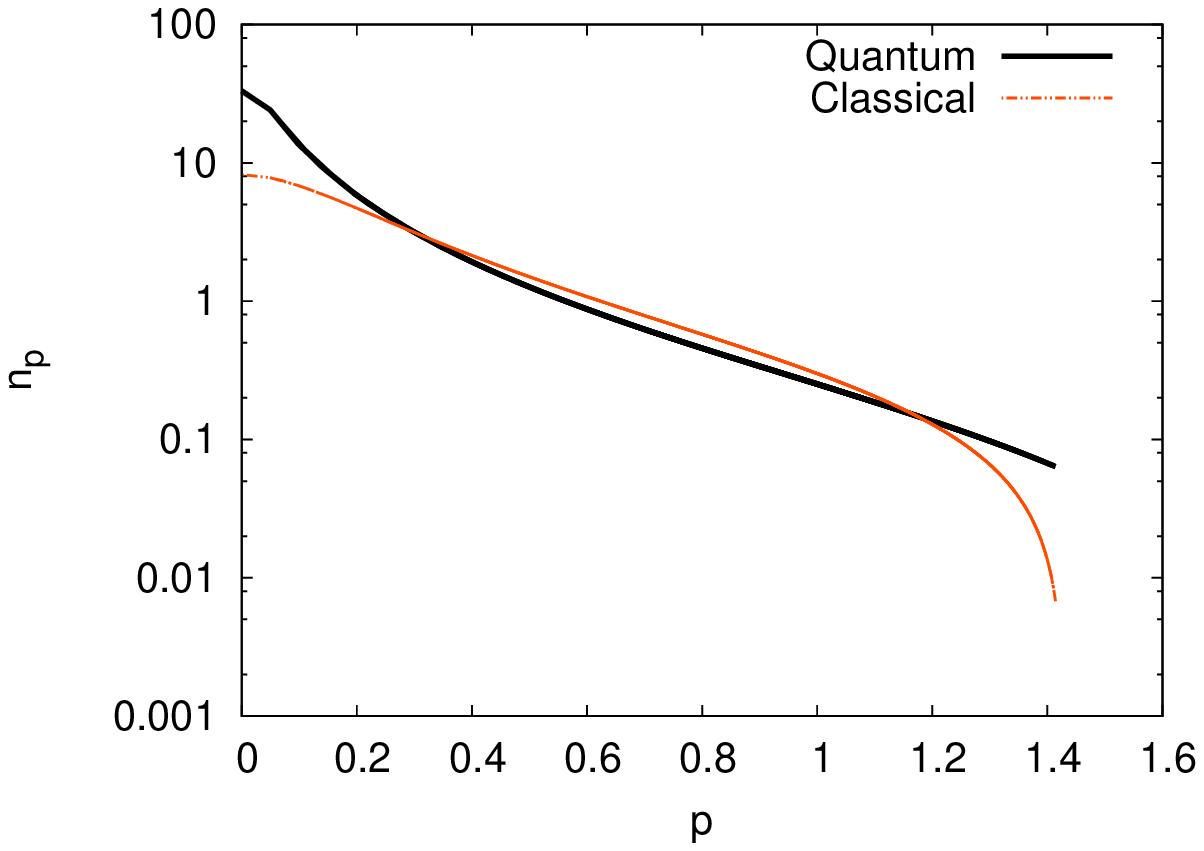}
\end{tabular}
\caption{The distribution $n_p$   defined in (\ref{freean}) at strong coupling $\lambda = 10$ without expansion. Left: $t=50t_0$; Right: $t=150t_0$. \label{n2}  }
\end{figure}

 Similarly, in the non-expanding case we plot in Fig.~\ref{n2} the distribution $n_p$ 
 defined in (\ref{freean}) as a function of $p=\sqrt{p_T^2+p_z^2}$. Again the distribution in the 2PI case develops an exponential tail already at $t=50t_0$, but the classical distribution follows a power law and becomes negative.


\section{Conclusions}
In this paper, we have for the first time performed the 2PI simulation of the $O(N)$ model with longitudinal expansion. This model is unique in that, to NLO in the $1/N$ expansion, one can  resum the 2PI diagrams to all orders in the coupling constant. Therefore, it is valid for all values of the coupling and also in the presence of violent instabilities.
  At weak coupling,  the parametric resonance is strong and the classical statistical approach  is an extremely good approximation. This means, however,  that we cannot see any hint of the quantum equilibration  within the time range explored in this work. Just for a (naive) comparison, we note that in heavy--ion collisions one would choose $\tau_0\sim 1/Q_s\sim 0.1\,$fm (the inverse of the saturation momentum) and the lifetime of the QGP is $\lesssim 5\,$fm, so in practice we have a constraint $\tau/\tau_0 \lesssim 50$. Thus, in order to achieve the truly equilibrated system (not just the establishment of an equation of state) in such a short time, pictures based on the weak coupling limit may not be a good starting point. Instead, we have seen that in the 2PI simulation at strong coupling the difference from the classical statistical approximation does show up within a reasonable time scale, and plays a crucial role in developing  the Bose--Einstein (or the exponential) distribution
   in $p_T$.

\vspace{10mm}
\section*{Acknowledgments}

We thank  Yukawa Institute for Theoretical Physics for providing computational facilities essential for the completion of this work.

\appendix

\section{Free correlation function}
The quantization of the scalar field in the $\tau$--$\eta$ coordinates was studied in \cite{Makhlin:1996dt}. In (1+2)--dimensions, the canonical commutation relation
\beq
[\phi(\tau,x_T,\eta),\tau \partial_\tau \phi(\tau,x'_T,\eta')] = i\delta(\eta-\eta')\delta(x_T-x'_T)\,,
\eeq
admits the following mode expansion
\begin{align}
\phi(x) = \int dp_T d p_\eta
\left[
a_{p_T,p_\eta} \xi_{p_T,p_\eta} (x) +  a^\dagger _{p_T, p_\eta} \xi_{p_T,p_\eta}^* (x)
\right]\,,
\end{align}
Here the one--particle wavefunction
\beq
&& \quad \qquad \xi_{p_T,p_\eta} (x) =
\frac{e^{-\pi p_\eta /2}}{4 \pi^{1/2}}
H_{-i p_\eta}^{(2)} (m_T \tau) \,
e ^{- i p_\eta \eta- i p_T x_T }\,, \nonumber \\
&& \int \tau dx_T d\eta\,  \xi^*_{p'_\eta,p'_T}(x) \, i\frac{\overleftrightarrow{\partial}}{\partial \tau} \xi_{p_\eta,p_T}(x) = \delta(p_\eta -p'_\eta)\delta(p_T - p'_T)\,,
\eeq
with $m_T=\sqrt{p_T^2+m^2}$, is normalized such that the commutation relation of $a$ and $a^\dagger$ takes the form
\beq
 [a_{p_T,p_\eta}, a^\dagger_{p'_T,p'_\eta}]
 = \delta(p_T -p'_T)\delta (p_\eta-p'_\eta)\,.
 \eeq
The  statistical and the spectral  functions of the free theory are easily calculated
\begin{align}
 F(\tau,\tau',p)= \frac{1}{2}\langle 0 | \{\phi(\tau,p), \phi  (\tau',-p) \}  |0 \rangle =
 \frac{\pi}{4} Re \left(H^{(2)}_{ip_\eta}(m_T \tau) H^{(1)}_{ip_\eta}(m_T \tau') \right)\,,
 \end{align}
\begin{align}
\rho(\tau,\tau',p) = i\langle 0| [\phi(\tau,p),\phi(\tau',-p)] |0\rangle=-\frac{\pi}{2} Im \left(H^{(2)}_{ip_\eta}(m_T \tau) H^{(1)}_{ip_\eta}(m_T \tau')  \right)\,.
 \end{align}

The approximate form at early times $\tau,\tau' \to 0$ is
\beq
 F(\tau,\tau,p) &\approx& \frac{1}{2p_\eta \tanh p_\eta \pi} +\frac{\cos\left(2p_\eta \ln \frac{m_{T} \,\tau}{2} -2\theta(p_\eta)\right)}{2p_\eta \sinh p_\eta \pi}\,, \label{finite} \\
 \rho(\tau,\tau',p) &\approx& -\frac{1}{p_\eta} \sin \left(p_\eta \ln \frac{\tau'}{\tau}\right)\,,
 \eeq
 where
 \beq
 \theta(p_\eta)= \arg \left( \Gamma (ip_\eta) \right)\,,
 \eeq
  is the phase of the gamma function.
The late--time behavior  $\tau,\tau'\to \infty$ is
\beq
F(\tau,\tau',p) &\approx& \frac{1}{2m_T \sqrt{\tau \tau'}} \cos m_T (\tau-\tau')\,,
\label{asy}
 \\
\rho(\tau,\tau',p) &\approx& \frac{1}{m_T \sqrt{\tau\tau'} }\sin m_T (\tau-\tau')\,.
\label{asy2}
\eeq
These are similar in form to the corresponding functions in the flat coordinates
\beq
F(t,t',p)=\frac{1}{2\omega_p}\cos \omega_p (t-t')\,, \label{free} \qquad
\rho(t,t',p)=\frac{1}{\omega_p} \sin \omega_p (t-t')\,,
\eeq
 where $\omega_p=\sqrt{p^2+m^2}$.
Finally we  note the first subleading term at $\tau = \tau' \,(\to \infty)$
\beq
F(\tau,\tau,p) \approx  \frac{1}{m_T \tau} \left(\frac{1}{2}-\frac{p_\eta^2+\frac{1}{4}}{4m_T^2 \tau^2}\right)\,. \label{correction}
\eeq

\end{document}